\documentstyle[12pt]{article}

         \def\la{\lambda}
         \def\be{\begin{equation}}
         \def\bea{\begin{eqnarray}}

         \def\ee{\end{equation}}
         \def\eea{\end{eqnarray}}
         \def\R{\rm {I\kern-.200em R}}
         \def\C{\rm {I\kern-.520em C}}
         \def\c{\chi}
         \def\ba{\begin{array}}
         \def\ea{\end{array}}
         \def\be{\begin{equation}}
         \def\bea{\begin{eqnarray}}
         \def\eea{\end{eqnarray}}
         \def\ee{\end{equation}}
         \def\tr{{\rm tr}}
         \def\p{{\rm P}}
         
\begin{document}
\begin{titlepage}
\hfill
\vbox{
    \halign{#\hfil         \cr
            hep-th/9612018      \cr
           } 
      }  
\vskip 10 mm
\begin{center}
{ \Large \bf
        Observables of the generalized 2D Yang--Mills theories}
{ \Large \bf
        on arbitrary surfaces: a path integral approach}

\vskip 15 mm
{ \bf M. Khorrami$^{a,b,c}$\footnote{e-mail:mamwad@netware2.ipm.ac.ir},
M. Alimohammadi$^{a,b}$\footnote {e-mail:alimohmd@@netware2.ipm.ac.ir}}
\vskip 10 mm
{\it
 $^a$ Department of Physics, Tehran University,
             North-Kargar Ave.\\
Tehran, Iran \\
 $^b$ Institute for Studies in Theoretical Physics and Mathematics,\\
           P.O.Box  5531, Tehran 19395, Iran\\
 $^c$ Institute for Advanced Studies in Basic Sciences,
 P.O.Box 159,\\
Gava Zang, Zanjan 45195, Iran\\

  }
\end{center}
\vskip 3cm
\begin{abstract}
Using the path integral method, we calculate the partition function and the
generating functional (of the field strengths) of the generalized 2D
Yang-Mills theories in the Schwinger--Fock gauge. Our calculation is done
for arbitrary 2D orientable, and also nonorientable surfaces.
\end{abstract}

\end{titlepage}
\newpage
\section{\bf Introduction}
The 2D Yang--Mills theory (YM$_2$) is a theoretical laboratory for
understanding the main theory of particle physics, QCD$_4$. It is
well known that YM$_2$ is a solvable model, and in recent years there have
been much effort to analyse the different aspects of this theory. A lattice
formulation of YM$_2$ has been known for a long time \cite{1}, and the
physical quantities, like the partition function and the expectation
values of the Wilson loops, have been calculated in this context for
arbitrary compact Riemann surfaces \cite{2,3}. The continuum (path integral)
approach has also been studied in \cite{4} and \cite{5}, and the Green
functions of field strengths have been calculated in [6-8].

In recent years, an interesting generalization of the 2D Yang-Mills theories
has been introduced. In fact, as a 2D counterpart of the theory of strong
interactions, pure YM$_2$ is not unique, and it is possible to generalize it
without losing properties such as invariance under area-preserving
diffeomorphisms and lack of propagating degrees of freedom.
In ref.\cite{W} , these generalized model have been introduced in the framework
of the BF theory, and the partition function of these theories have been
obtained by regarding the general Yang-Mills action as perturbation of the
topological theory at zero area.In \cite{9}, the large $N$-limit of such theories
(called generalized 2D Yang--Mills theories (gYM$_2$'s) in \cite{9} ),when
coupled to fundamental fermions of $SU(N)$ have been studied. And the
authors of \cite{10} have generalized the Migdal's suggestion about the
local factor of plaquettes, and have shown that this generalization
satisfies the necessary requirements. In this way they have found
the partition function and the
expectation values of Wilson loops of gYM$_2$'s. For a reviw see \cite{11}.

In this paper we are going to handle gYM$_2$ by the standard path integral
approach. In section 2, we caculate the partition function, the wave
functions, and the expectation values of Wilson loops of gYM$_2$, by doing
the path integration of the corresponding BF theory in the Schwinger--Fock
gauge. Our calculation is done for arbitrary orientable, and also
nonorientable surfaces. It is then seen that our results are in agreement
with those of \cite{10}. In section 3, we investigate the question of the
correlation functions of field strengths of gYM$_2$. For this purpose, we
calculate the generating functional of these fields, and show that our
general results reduce to the known results of YM$_2$ \cite{7,8} in the
special case.

\section{\bf The wave functions of gYM$_2$}
The gYM$_2$ is defined by
\be\label{1} \exp{S(\xi )}:=\int DB\;\exp[{\cal S}(B,\xi )],\ee
where
\be\label{2} {\cal S}(B,\xi ):=\int [-i\tr (B\xi )+f(B)]d\mu .\ee
In this equation, $\xi (x)$ is defined through
\be \epsilon_{\mu\nu}\xi (x):=F_{\mu\nu}(x),\ee
$\xi$ and $B$ are elements of the adjoint representation of the Lie algebra
corresponding to the gauge group $G$, $f$ is an arbitrary class function of
$B$:
\be\label{4} f(UBU^{-1})=f(B),\qquad\forall U\in G,\ee
and
\be d\mu :={1\over 2}\epsilon_{\mu\nu}d x^\mu\wedge d x^\nu.\ee
If $f(B)$ is a quadratic function, $f(B)=\tr B^2$, it is easy to do the
integration (\ref{1}), and see that it is the standard YM$_2$.

To begin, we calculate the wave function corresponding to (\ref{1}) on a
disk $D$ with the boundary condition $\p\exp\oint_{\gamma =\partial D}
A=g\in G$:
\be \psi_D(g)=\int D\xi\; DB\;\exp [{\cal S}(B,\xi )]\delta\left(
\p\exp\oint_{\gamma}A,g\right) .\ee
One can expand the (class) delta function in terms of the characters of
the irreducible representations of G:
\be\delta\left(\p\exp\oint_{\gamma}A, g\right) =\sum_\lambda\chi_\lambda
(g^{-1})\chi_\lambda\left(\p\exp\oint_{\gamma}A\right) ,\ee
and use the following fermionic
path integral representation of the Wilson loops \cite{4,12}:
\be\chi_\lambda\left(\p\exp\oint_{\gamma =\partial D}A\right) =
\int D\eta\; D\bar\eta\;\exp\left[\int_0^1\bar\eta (t)\dot\eta (t)dt+
\oint_\gamma\bar\eta A_{(\lambda )}\eta\right]\eta^\alpha(0)
\bar\eta_\alpha (1).\ee
Here $\eta$ and $\bar\eta$ are Grassmann valued vectors in the
representation $\lambda$. In the Shwinger--Fock gauge, we also have
\be A_\mu^a(x)=\int_0^1 dr\; rx^\nu F_{\nu\mu}^a(rx),\ee
and
\be F=dA.\ee
So
\bea \psi_D(g)&=&\sum_\lambda\chi_\lambda (g^{-1)}\int D\eta\; D\bar\eta\;
DB\; D\xi\;\exp\left[\int_0^1\bar\eta (t)\dot\eta (t)dt\right]
\eta^\alpha (0)\bar\eta_\alpha (1)\cr &&\times\exp\left\{\int_D[-i\tr (B\xi)
+f(B)+\bar\eta (t)T_a(\lambda )\eta (t)\xi^a(x)]d\mu\right\} ,\eea
where we have parametrized the disk by the angle ($t$) and radius ($r$)
variables. $T_a(\lambda )$'s are the generators of $G$ in the representation
$\lambda$. An integration over $\xi$, and then $B$, yields
\be \psi_D(g)=\sum_\lambda\chi_\lambda (g^{-1)}\int D\eta\; D\bar\eta\;
\exp\left\{\int_0^1\bar\eta (t)\dot\eta (t)dt +\int_D
f[i\bar\eta T(\lambda )\eta ]d\mu\right\} \eta^\alpha (0)\bar
\eta_\alpha (1),\ee
where
\be T(\lambda):=T_a(\lambda )\otimes T^a,\ee
and
\be \bar\eta T(\lambda )\eta =\bar\eta T_a(\lambda )\eta T^a,\ee
in which $T^a$'s are the generators of $G$ in the adjoint representation.
Now, using the fermionic propagator in one dimension:
\be\int D\eta\; D\bar\eta\exp\left[\int_0^1\bar\eta (t)\dot\eta (t)dt\right]
\eta^\alpha (t')\bar\eta_\beta (t'')=\delta^\alpha_\beta\theta (t''-t'),\ee
it is seen that
\be\label{13}\psi_D(g)=\sum_\lambda\chi_\lambda (g^{-1})\tr\;
\exp\{ A(D)f[iT(\lambda)]\} .\ee
Note that $f$ can be expanded in terms of the components of $B$. The
coefficients of this expansion are symmetric tensors, so that no ambiguity
arises in (\ref{13}) because of the noncommutativity of $T_a$'s.

Now, from (\ref{4}) we have
\be f[i\; 1\otimes U\; T(\lambda )\; 1\otimes U^{-1}]=f[iT(\lambda )].\ee
However,
\be 1\otimes U\; T(\lambda )\; 1\otimes U^{-1}=
    U(\lambda )\otimes 1\; T(\lambda )\; U^{-1}(\lambda )\otimes 1,\ee
and
\be f[i\; U(\lambda )\otimes 1\; T(\lambda )\; U^{-1}(\lambda )\otimes 1]=
    U(\lambda )f[iT(\lambda )]U^{-1}(\lambda ),\ee
which shows that
\be U(\lambda )f[iT(\lambda )]U^{-1}(\lambda )=f[iT(\lambda )],\qquad
\forall U\in G.\ee
This means that $f[iT(\lambda )]$ is proportional to identity:
\be f[iT(\lambda )]=f_\lambda 1_\lambda.\ee
So, (\ref{13}) becomes
\be\psi_D(g)=\sum_\lambda\chi_\lambda (g^{-1})d_\lambda
\exp [A(D)f_\lambda ].\ee
This is the same result of \cite{10}.
Now, we proceed by the same procedure as was followed in YM$_2$ \cite{4,8}
and calculate the partition function of gYM$_2$ on arbitrary (orientable or
nonorientable) surfaces: We glue the wave functions in a suitable way (for
detailes see \cite{8}). The final result is
\be\label{21}\psi_{\Sigma_{g, s, r}}(g_1, \cdots , g_n)=\sum_\lambda
h_\lambda^{r+2s}d_\lambda^{2-2g-2s-r-n}\chi_\lambda (g_1^{-1}) \cdots
\chi_\lambda (g_n^{-1})\exp \left[ A(\Sigma_{g, s, r})f_\lambda\right] .\ee
$\Sigma_{g, s, r}$ is a connected sum of an orientable surface of genus $g$
with $s$ Klein bottles and $r$ projective planes, having $n$ boundaries
$\gamma_1, \cdots , \gamma_n$, with boundary conditions
$\p\exp\oint_{\gamma_i} A=g_i\in G$. $h_\lambda$ is defined through
\be h_\lambda :=\int\chi_\lambda (g^2)dg;\ee
it is zero unless the representation $\lambda$ is self conjugate. If so,
this representation has an invariant bilinear form. Then, $f_\lambda =1$ if
this form is symmetric, and $f_\lambda =-1$ if it is antisymmetric
\cite{13}.

In the case of orientable surfaces ($r=s=0$), our general result (\ref{21})
coincides with that of \cite{10}. The expectation values of Wilson loops are
also found by the same method followed in YM$_2$, and leads to the same
result of YM$_2$, except that the second Casimir must be replaced by
$f_\lambda$.

\section{\bf The generating functinal $Z[J]$ of gYM$_2$}

To calculate the Green functions of the field stength $\xi^a$'s , we again begin
with the disk and calculate the wavefunction of gYM$_2$ on a disk $D$ , with a
source term coupled to $\xi$ :
\be \psi_D[J]=\int D\xi\; DB\;\exp\left\{\int [-i {\rm tr} (B\xi )+f(B)+\xi^a
J_a] d\mu\right\}
\delta({\rm Pexp}\oint_{\gamma =\partial D} A,g). \ee
Following the same steps of the previous section , we arrive at :
\be \psi_D[J]=\sum_\la \c_\la (g^{-1})\int D\eta\; D{\bar \eta}\;
\exp\left\{\int f[iJ^a(x)+i{\bar \eta}(t)T^a
\eta  (t)]d\mu\right\}\exp\left[\int_0^1 dt {\bar \eta}(t)
{\dot \eta}(t)\right]\eta^\alpha(0){\bar \eta }_\alpha(1) . \ee
Again we can expand the first exponential and calculate the resulting Green
functions of the free-fermionic theory .Using the fact that the tensorial
coefficients in $f$ are totally symmetric , we finally find :
\be \psi_D[J]=\sum_\la \c_\la (g^{-1}){\rm tr }_\lambda[{\cal P} {\rm exp}\int f(iJ^a
(x)+iT^a)d\mu].\ee
In the above equation ${\cal P}$ stands for ordering according to the angle
variable on the disk.

As an example , consider the YM$_2$ , in which $f(B)=- \epsilon {\rm tr}(B^2)$ .
In this case (27) reduces to :
\be \psi_D[J]=\exp(\epsilon \int J^aJ_a d\mu)\sum_\la \c_\la (g^{-1})
\exp\left[-\epsilon c_2(\la ) A(D)\right]{\rm tr }_\lambda[{\cal P} {\rm exp}2\epsilon
\int dt \int dr {\sqrt g}J(r,t)],\ee
which is in agrrement with the result obtained in [7].

Now if we glue (27) to the wavefunction (23) (with $n=1$ ) :
\be Z_{\Sigma_{g,s,r}}[J]=\int dg \psi_D[J,g]\psi(\Sigma_{g,s,r},g^{-1}) , \ee
we find the following generating functional $Z[J]$ of gYM$_2$ on $\Sigma_{g,s,r}$ :
\be Z_{\Sigma_{g,s,r}}[J]=\sum_\la h_\la^{r+2s}d(\la)^{2-2g-2s-r-1}
\exp\left[ A(\Sigma_{g,s,r})f_\la\right]
{\rm tr }_\lambda\left\{{\cal P} {\rm exp}\int f\left[iJ^a(x)+iT^a\right]d\mu
\right\} .\ee
Functinal differentiating of (30) with respect to $J(x)$ gives us the Green
functions of $\xi$'s in the Schwinger-Fock gauge.

\vskip 1cm
{\bf Acknowledgement} We would like to thank the research vice-chancellor
of Tehran University, this work was partially supported by them.

\end{document}